\documentclass[11pt]{cernrep}
\usepackage{graphicx}
\usepackage{cite}
\usepackage{epsfig}
\begin{document}

\title{\vspace*{-.3in}
\flushright{\rm \small UCD-2004-16 \\[-2pt]IFIC/04-03\\[-2pt]LPT-Orsay-04-12\\[-2pt] SHEP-03-41\\[-8pt] hep-ph/0401228} \\
 \centerline{NMSSM Higgs Discovery at the LHC~\footnote{~~~~To appear in the Proceedings of the Les Houches Workshop 2003: ``Physics at TeV Colliders'', ed. F. Boudjema}}}
\author{U. Ellwanger$^1$, J.F. Gunion$^2$, C. Hugonie$^3$ and S. Moretti$^4$}
\institute{$^1$~LPTHE, Universit\'e de Paris XI, B\^atiment 210, F091405 Orsay Cedex, France; $^2$~Department of Physics, U.C. Davis, Davis, CA 95616; $^3$~AHEP Group,
I. de F\'isica Corpuscular -- CSIC/Universitat de
Val\`encia, Edificio Institutos de Investigaci\'on,
Apartado de Correos 22085, E-46071, Valencia, Spain;
$^4$~School of Physics, Department of Physics and Astronomy, University of Southampton, Southampton, SO171BJ, UK}

\maketitle

\vspace{1em}
\def\cnone{\widetilde \chi_1^0}
\def\cpone{\widetilde \chi_1^+}
\def\cmone{\widetilde \chi_1^-}
\def\mcnone{m_{\cnone}}
\def\mcpmone{m_{\widetilde\chi_1^{\pm}}}
\def\tev{~{\rm TeV}}
\def\gev{~{\rm GeV}}
\def\mgut{M_{U}}
\def\what{\widehat}
\def\hpm{H^{\pm}}
\def\ie{{\it i.e.}}
\def\anti{\overline}
\def\br{BR}
\def\gam{\gamma}

\vspace*{-.2in}
\centerline{\bf Abstract}

\noindent 
We demonstrate that Higgs discovery at the LHC is possible
in the context of the NMSSM even for those scenarios such
that the only strongly produced Higgs boson is a very SM-like
CP-even scalar
which decays almost entirely 
to a pair of relatvely light CP-odd states. 
In combination with other search channels, we are on
the verge of demonstrating that detection of at least
one of the NMSSM Higgs bosons is guaranteed at the LHC
for accumulated luminosity of $300~{\rm fb}^{-1}$.

\vspace*{-.15in}
\section{Introduction}

One of the most attractive supersymmetric models
is the Next to Minimal Supersymmetric Standard
Model (NMSSM) (see
\cite{Ellis:1989er,Ellwanger:2001iw} and
references therein) which extends the MSSM by the
introduction of just one singlet superfield, $\what
S$. When the scalar component of $\what S$
acquires a TeV scale vacuum expectation value (a
very natural result in the context of the model),
the superpotential term $\what S \what H_u \what
H_d$ generates an effective $\mu\what H_u \what
H_d$ interaction for the Higgs doublet
superfields.  Such a term is essential for
acceptable phenomenology. No other SUSY model
generates this crucial component of the
superpotential in as natural a fashion. Thus, the
phenomenological implications of the NMSSM at
future accelerators should be considered very
seriously.  One aspect of this is the fact that
the $h,H,A,\hpm$ Higgs sector of the MSSM is
extended so that there are three CP-even Higgs
bosons ($h_{1,2,3}$, $m_{h_1}<m_{h_2}<m_{h_3}$),
two CP-odd Higgs bosons ($a_{1,2}$,
$m_{a_1}<m_{a_2}$) (we assume that CP is not
violated in the Higgs sector) and a charged Higgs
pair ($h^\pm$). An important question is then the
extent to which the no-lose theorem for MSSM Higgs
boson discovery at the LHC (after LEP constraints)
is retained when going to the NMSSM; \ie\ is the
LHC guaranteed to find at least one of the
$h_{1,2,3}$, $a_{1,2}$, $h^\pm$? The first exploration
of this issue appeared in \cite{Gunion:1996fb},
with the conclusion that for substantial portions
of parameter space the LHC would be unable to
detect any of the NMSSM Higgs bosons.
Since then, there have been improvements in many
of the detection modes and the addition of
new ones. These will be summarized below
and the implications reviewed.  However, these
improvements and additions do not address the 
possibly important $h\to aa$ type decays that could
suppress all other types of signals \cite{Gunion:1996fb,Dobrescu:2000jt}.

One of the key ingredients in the no-lose theorem
for MSSM Higgs boson discovery is the fact that
relations among the Higgs boson masses are such
that decays of the SM-like Higgs boson to $AA$ are
only possible if $m_A$ is quite small, a region
that is ruled out by LEP by virtue of the fact
that $Z\to hA$ pair production was not detected
despite the fact that the relevant coupling is
large for small $m_A$.  In the NMSSM, the lighter
Higgs bosons, $h_1$ or $h_2$, can be SM-like (in
particular being the only Higgs with substantial
$WW/ZZ$ coupling) without the $a_1$ necessarily
being heavy.  In addition, this situation is not
excluded by LEP searches for $e^+e^-\to Z^*\to
h_{1,2}a_1$ since, in the NMSSM, the $a_1$ can
have small $Zh_2 a_1$ ($Zh_1 a_1$) coupling when
$h_1$ ($h_2$) is SM-like. [In addition, sum rules
require that the $Zh_1 a_1$ ($Zh_2 a_1$) coupling
is small when the $h_1WW$ ($h_2WW$) couplings are
near SM strength.]  As a result, NMSSM parameters that are
not excluded by current data can be chosen so that
the $h_{1,2}$ masses are moderate in size ($\sim
100-130$~GeV) and the $h_1\to a_1a_1$ or $h_2\to
a_1a_1$ decays are dominant.  Dominance of such
decays falls outside the scope of the usual
detection modes for the SM-like MSSM $h$ on which
the MSSM no-lose LHC theorem largely relies.

In Ref.~\cite{Ellwanger:2001iw}, a partial no-lose
theorem for NMSSM Higgs boson discovery at the LHC
was established.  In particular, it was shown that
the LHC would be able to detect at least one of
the Higgs bosons (typically, one of the lighter
CP-even Higgs states) throughout the full
parameter space of the model, excluding only those
parameter choices for which there is sensitivity
to the model-dependent decays of Higgs bosons to
other Higgs bosons and/or superparticles.  Here,
we will address the question of whether or not
this no-lose theorem can be extended to those
regions of NMSSM parameter space for which Higgs
bosons can decay to other Higgs bosons.  We find
that the parameter choices such that the
``standard'' discovery modes fail {\it would}
allow Higgs boson discovery if detection of $h\to
aa$ decays is possible. (When used generically,
the symbol $h$ will now refer to $h=h_1$, $h_2$ or
$h_3$ and the symbol $a$ will refer to $a=a_1$ or
$a_2$).  Detection of $h\to aa$ will be difficult
since each $a$ will decay primarily to 
$b\anti b$ (or 2 jets if $m_a<2m_b$),
$\tau^+\tau^-$, and, possibly, $\cnone\cnone$,
yielding final states that will typically have
large backgrounds at the LHC.

In \cite{Ellwanger:2001iw} we scanned the
parameter space, removing parameter choices ruled
out by constraints from LEP on Higgs boson
production, $e^+ e^- \to Z h$ or $e^+ e^- \to h a$
\cite{LEPLEPHA}, and eliminating parameter choices
for which one Higgs boson can decay to two other
Higgs bosons or a vector boson plus a Higgs boson.
For the surviving regions of parameter space, we
estimated the statistical significances
($N_{SD}=S/\sqrt B$) for all Higgs boson detection
modes so far studied at the LHC \cite{CMS,ATLAS,Zeppenfeld:2000td,Zeppenfeld:2002ng}.
%,ATLAS,6.2r,Zeppenfeld:2002ng}. 
These are (with $\ell=e,\mu$)

1) $g g \to h/a \to \gamma \gamma$;\par
2) associated $W h/a$ or $t \bar{t} h/a$ production with 
$\gamma \gamma\ell^{\pm}$ in the final state;\par
3) associated $t \bar{t} h/a$ production with $h/a \to b \bar{b}$;\par
4) associated $b \bar{b} h/a$ production with $h/a \to \tau^+\tau^-$;\par
5) $g g \to h \to Z Z^{(*)} \to$ 4 leptons;\par
6) $g g \to h \to W W^{(*)} \to \ell^+ \ell^- \nu \bar{\nu}$;\par
7) $W W \to h \to \tau^+ \tau^-$;\par
8) $W W \to h\to W W^{(*)}$.\par

\noindent
For an integrated luminosity of $300~{\rm
  fb}^{-1}$ at the LHC, all the surviving points
yielded $N_{SD}>10$ after combining all modes,
including the $W$-fusion modes. Thus, NMSSM Higgs
boson discovery by just one detector with
$L=300~{\rm fb}^{-1}$ is essentially guaranteed
for those portions of parameter space for which
Higgs boson decays to other Higgs bosons or
supersymmetric particles are kinematically
forbidden.

In this work, we investigate the complementary
part of the parameter space, where {\it at least
  one} Higgs boson decays to other Higgs bosons.
To be more precise, we require at least one of the
following decay modes to be kinematically allowed:
\begin{eqnarray}
& i) \ h \to h' h' \; , \quad ii) \ h \to a a \; , \quad iii) \ h \to h^\pm
h^\mp \; , \quad iv) \ h \to a Z \; , \nonumber \\
& v) \ h \to h^\pm W^\mp \; , \quad vi) \ a' \to h a \; , \quad vii) \ a \to h
Z \; , \quad viii) \ a \to h^\pm W^\mp \; .
\end{eqnarray}
After searching those regions of parameter space
for which one or more of the decays $ i) - viii)$
is allowed, we found that the only subregions for
which discovery of a Higgs boson in modes 1) -- 8)
was not possible correspond to NMSSM parameter
choices for which (a) there is a light CP-even
Higgs boson with substantial doublet content that
decays mainly to two still lighter CP-odd Higgs
states, $h\to aa$, and (b) all the other Higgs
states are either dominantly singlet-like,
implying highly suppressed production rates, or
relatively heavy, decaying to $t\anti t$, to one
of the ``difficult'' modes $i) - viii)$ or to a
pair of sparticles. In such cases, the best
opportunity for detecting at least one of the
NMSSM Higgs bosons is to employ $WW\to h$
production and develop techniques for extracting a
signal for the $h\to aa\to jj\tau^+\tau^-$ 
(including $jj=b\anti b$) process.  We
have performed a detailed simulation of 
the $aa\to jj \tau^+\tau^-$ 
final state and find that its detection may be possible
after accumulating $300~{\rm fb}^{-1}$ in both the
ATLAS and CMS detectors.  

\vspace*{-.15in}
\section{The model and scanning procedures}

We consider the simplest version of the NMSSM
\cite{Ellis:1989er}, where the term $\mu \widehat
H_1 \widehat H_2$ in the superpotential of the
MSSM is replaced by (we use the notation $\widehat
A$ for the superfield and $A$ for its scalar
component field)
\begin{equation}\label{2.1r}
\lambda \widehat H_1 \widehat H_2 \widehat S\ + \ \frac{\kappa}{3} \widehat S^3
\ \ ,
\end{equation}
\noindent so that the superpotential is scale invariant. 
We make no assumption on ``universal'' soft terms.
Hence, the five soft supersymmetry breaking terms
\begin{equation}\label{2.2r}
m_{H_1}^2 H_1^2\ +\ m_{H_2}^2 H_2^2\ +\ m_S^2 S^2\ +\ \lambda
A_{\lambda}H_1 H_2 S\ +\ \frac{\kappa}{3} A_{\kappa}S^3
\end{equation}
\noindent are considered as independent. 
The masses and/or couplings of sparticles will
be such that their contributions to the
loop diagrams inducing Higgs boson production by
gluon fusion and Higgs boson decay into $\gamma
\gamma$ are negligible. 
In the gaugino sector, we chose $M_2=1\tev$ (at low scales).
Assuming universal gaugino masses at the coupling
constant unification scale,
this yields $M_1\sim 500\gev$ and $M_3\sim 3\tev$.
In the squark sector, as particularly relevant
for the top squarks which
appear in the radiative corrections to the Higgs
potential, we chose the soft masses $m_Q = m_T
\equiv M_{susy}= 1$ TeV, and varied the stop
mixing parameter
\begin{equation}\label{2.4r}
X_t \equiv 2 \ \frac{A_t^2}{M_{susy}^2+m_t^2} \left ( 1 -
\frac{A_t^2}{12(M_{susy}^2+m_t^2)} \right ) \ .
\end{equation} 
\noindent As in the MSSM, 
the value $X_t = \sqrt{6}$ -- so called maximal
mixing -- maximizes the radiative corrections to
the Higgs boson masses, and we found that it leads
to the most challenging points in the parameter
space of the NMSSM.  We adopt the convention
$\lambda,\kappa > 0$, in which $\tan\beta$ can
have either sign. We require $|\mu_{\rm eff}|\ >\ 
100$~GeV; otherwise a light chargino would have
been detected at LEP. The only possibly light SUSY particle
will be the $\cnone$.  A light $\cnone$ is a frequent
characteristic of parameter choices that yield a 
light $a_1$.

We have performed a numerical scan over the free
parameters.  For each point, we computed the
masses and mixings of the CP-even and CP-odd Higgs
bosons, $h_i$ ($i=1,2,3$) and $a_j$ ($j=1,2$),
taking into account radiative corrections up to
the dominant two loop terms, as described in
\cite{Ellwanger:1999ji}.  We eliminated parameter
choices excluded by LEP
constraints~\cite{LEPLEPHA} on $e^+ e^- \to Z h_i$
and $e^+ e^- \to h_i a_j$. The latter provides an
upper bound on the $Zh_ia_j$ reduced coupling,
$R'_{ij}$, as a function of $m_{h_i}+m_{a_j}$ for
$m_{h_i} \simeq m_{a_j}$.  Finally, we calculated
$m_{h^\pm}$ and required $m_{h^\pm} > 155$~GeV, so
that $t \to h^\pm b$ would not be seen.

In order to probe the complementary part of the
parameter space as compared to the scanning of
Ref. \cite{Ellwanger:2001iw}, we required that at
least one of the decay modes $i) - viii)$ is
allowed.  For each Higgs state, we calculated all
branching ratios including those for modes $i) -
viii)$, using an adapted version of the FORTRAN
code HDECAY \cite{Djouadi:1998yw}. We then
estimated the expected statistical significances
at the LHC in all Higgs boson detection modes 1)
-- 8) by rescaling results for the SM Higgs boson
and/or the MSSM $h, H$ and/or $A$. The
rescaling factors are determined by $R_i$, $t_i$
and $b_i=\tau_i$, the ratios of the $VVh_i$,
$t\anti t h_i$ and $b\anti b h_i,\tau^+\tau^- h_i$
couplings, respectively, to those of a SM Higgs
boson.  Of course $|R_i| < 1$, but $t_i$ and $b_i$
can be larger, smaller or even differ in sign with
respect to the SM. For the CP-odd Higgs bosons,
$R_i'=0$ at tree-level; $t'_j$ and $b'_j$ are the
ratios of the $i\gamma_5$ couplings for $t\bar{t}$
and $b\bar{b}$, respectively, relative to SM-like
strength.  A detailed discussion of the procedures
for rescaling SM and MSSM simulation results for
the statistical significances in channels 1) -- 8)
will appear elsewhere.

\begin{table}[p]
\begin{center}
\footnotesize
\vspace*{-.2in}
\hspace*{-.5in}
\begin{tabular} {|l|l|l|l|l|l|l|} 
\hline
Point Number & 1 & 2 & 3 & 4 & 5 & 6  \\
\hline \hline
Bare Parameters &\multicolumn{6}{c|}{} \\
\hline
$\lambda$            & 0.2872 & 0.2124 & 0.3373 & 0.3340 & 0.4744 & 0.5212 \\
\hline
$\kappa$             & 0.5332 & 0.5647 & 0.5204 & 0.0574 & 0.0844 & 0.0010 \\
\hline
$\tan\beta$          &   2.5  &   3.5  &   5.5  &    2.5 &    2.5 & 2.5 \\
\hline
$\mu_{\rm eff}~({\rm GeV})$&    200 &    200 &    200 &    200 &    200 & 200 \\
\hline
$A_{\lambda}~({\rm GeV})$  &    100 &      0 &     50 &    500 &    500 & 500 \\
\hline
$A_{\kappa}~({\rm GeV})$   &      0 &      0 &      0 &      0 &      0 & 0 \\
\hline \hline
CP-even Higgs Boson Masses and Couplings &\multicolumn{6}{c|}{} \\
\hline \hline
$m_{h_1}$~(GeV)      &    115 &    119 &    123 &     76 &     85 &  51\\
\hline
$R_1 $               &   1.00 &   1.00 &  -1.00 &   0.08 &   0.10 &  -0.25\\
\hline
$t_1 $               &   0.99 &   1.00 &  -1.00 &   0.05 &   0.06 &  -0.29\\
\hline
$b_1 $               &   1.06 &   1.05 &  -1.03 &   0.27 &   0.37 &  0.01\\
\hline
Relative 
gg Production Rate   &   0.97 &   0.99 &   0.99 &   0.00 &   0.01 &  0.08\\
\hline
$\br(h_1\to 
b\anti b)$           &   0.02 &   0.01 &   0.01 &   0.91 &   0.91 &  0.00\\
\hline
$\br(h_1\to 
\tau^+\tau^-)$      &   0.00 &   0.00 &   0.00 &   0.08 &   0.08 &  0.00\\
\hline
$\br(h_1\to a_1 a_1)$&   0.98 &   0.99 &   0.98 &   0.00 &   0.00 &  1.00\\
\hline \hline

$m_{h_2}$~(GeV)      &    516 &    626 &    594 &    118 &    124 &  130\\
\hline
$R_2 $               &  -0.03 &  -0.01 &   0.01 &  -1.00 &  -0.99 &  -0.97\\
\hline
$t_2 $               &  -0.43 &  -0.30 &  -0.10 &  -0.99 &  -0.99 &  -0.95\\
\hline
$b_2 $               &   2.46 &  -3.48 &   3.44 &  -1.03 &  -1.00 &  -1.07\\
\hline
Relative
gg Production Rate   &   0.18 &   0.09 &   0.01 &   0.98 &   0.99 &  0.90\\
\hline
$\br(h_2\to 
b\anti b)$           &   0.01 &   0.04 &   0.04 &   0.02 &   0.01 &  0.00\\
\hline
$\br(h_2\to 
\tau^+\tau^-)$      &   0.00 &   0.01 &   0.00 &   0.00 &   0.00 &  0.00\\
\hline
$\br(h_2\to a_1 a_1)$&   0.04 &   0.02 &   0.83 &   0.97 &   0.98 &  0.96\\
%\hline
%$\br(h_2\to h_1 h_1)$&   0.01 &   0.01 &   0.00 &   0.00 &   0.00 &  0.04\\
\hline \hline

$m_{h_3}$~(GeV)      &    745 &   1064 &    653 &    553 &    554 &  535\\
\hline \hline

CP-odd Higgs Boson Masses and Couplings &\multicolumn{6}{c|}{} \\
%and Couplings &\multicolumn{6}{c|}{} \\
\hline \hline
$m_{a_1}$~(GeV)      &     56 &      7 &     35 &     41 &     59 &  7\\
\hline
$t_1' $               &   0.05 &   0.03 &   0.01 &  -0.03 &  -0.05 &  -0.06\\
\hline
$b_1' $               &   0.29 &   0.34 &   0.44 &  -0.20 &  -0.29 &  -0.39\\
\hline
Relative
gg Production Rate   &   0.01 &   0.03 &   0.05 &   0.01 &   0.01 &  0.05\\
\hline
$\br(a_1\to 
b\anti b)$           &   0.92 &   0.00 &   0.93 &   0.92 &   0.92 &  0.00\\
\hline
$\br(a_1\to 
\tau^+\tau^-)$      &   0.08 &   0.94 &   0.07 &   0.07 &   0.08 &  0.90\\
\hline \hline

$m_{a_2}$~(GeV)      &    528 &    639 &    643 &    560 &    563 &  547\\
\hline 
Charged Higgs  
Mass (GeV)           &    528 &    640 &    643 &    561 &    559 &  539\\
\hline\hline
Most Visible of the LHC Processes 1)-8) &  2 ($h_1$) &  2 ($h_1$) &  8
                  ($h_1$) &  2 ($h_2$) &  8 ($h_2$)  &  8 ($h_2$)\\
\hline            
$N_{SD}=S/\sqrt B$ 
Significance of this process at $L=$300~${\rm fb}^{-1}$
                     &   0.48 &   0.26 &   0.55 &   0.62 &  0.53  & 0.16\\
\hline
\hline
$N_{SD}(L=300~{\rm fb}^{-1})$  for
$WW\to h\to aa\to jj \tau^+\tau^-$ at LHC & 50 &  22 &  69 &  63&  62 &  21 \\
\hline
\end{tabular}
\end{center}
\vspace*{-.2in}\caption{\label{tpoints}\footnotesize
Properties of selected scenarios that could escape detection
at the LHC. In the table, $R_i=g_{h_i VV}/g_{h_{SM} VV}$, 
$t_i=g_{h_i t\anti t}/g_{h_{SM} t\anti t}$ and $b_i=g_{h_ib\anti b}/g_{h_{SM} b\anti b}$ 
for $m_{h_{SM}}=m_{h_i}$; $t_1'$ and $b_1'$
are the $i\gam_5$ couplings of $a_1$ 
to $t\anti t$ and $b\anti b$ normalized
relative to the scalar 
$t\anti t$ and $b\anti b$ SM Higgs couplings.
We also give the $gg$ fusion production rate ratio,
$gg\to h_i/gg\to h_{SM}$, for $m_{h_{SM}}=m_{h_i}$. 
Important absolute branching
ratios are displayed. For points 2 and 6, the decays
$a_1\to jj$ ($j\neq b$) have 
$\br(a_1\to jj)\simeq 1-\br(a_1\to \tau^+\tau^-)$.
For the heavy $h_3$ and $a_2$, we give only their masses.
For all points 1 -- 6, the statistical
significances for the detection of any 
Higgs boson in any of the channels 1) --
8) are tiny; the next-to-last row gives their maximum 
together with the process number and 
the corresponding Higgs state.
The last row gives the statistical significance
of the new $WW\to h \to aa\to jj \tau^+\tau^-$
[$h=h_1$ ($h=h_2$) for points 1--3 (4--6)] LHC
signal explored here. 
}
\end{table}

In our set of randomly scanned points, we selected
those for which all the statistical significances
in modes 1) -- 8) are below $5\sigma$. We obtained
a lot of points, all with similar characteristics.
Namely, in the Higgs spectrum, we always have a
very SM-like CP-even Higgs boson with a mass
between 115 and 135~GeV ({\it i.e.} above the LEP
limit), which can be either $h_1$ or $h_2$, with a
reduced coupling to the gauge bosons $R_1 \simeq
1$ or $R_2\simeq 1$, respectively. This state
decays dominantly to a pair of (very) light CP-odd
states, $a_1a_1$, with $m_{a_1}$ between 5 and
65~GeV.  The singlet component of $a_1$ cannot be
dominant if we are to have a large $h_1 \to a_1
a_1$ or $h_2\to a_1a_1$ branching ratio when the
$h_1$ or $h_2$, respectively, is the SM-like Higgs
boson.  Further, when the $h_1$ or $h_2$ is very
SM-like, one has small $Zh_1a_1$ or $Zh_2a_1$ coupling,
respectively, so that $e^+ e^- \to
h_1 a_1$ or $e^+e^-\to h_2 a_1$ associated
production places no constraint on the light
CP-odd state at LEP. We have selected six
difficult benchmark points, displayed in
Table~\ref{tpoints}.  These are such that
$a_1\to\cnone\cnone$ decays are negligible or
forbidden.  (Techniques for cases such that 
$\cnone\cnone$ decay modes are important
are under development.)
For points 1 -- 3, $h_1$ is the
SM-like CP-even state, while for points 4 -- 6 it
is $h_2$. We have selected the points so that there is
some variation in the 
$h_{1,2}$ and $a_1$ masses. The
main characteristics of the benchmark points are
displayed in Table~\ref{tpoints}. Note the large
$\br(h\to a_1 a_1)$ of the SM-like $h$ ($h=h_1$
for points 1 -- 3 and $h=h_2$ for points 4 --6).
For points 4 -- 6, with $m_{h_1}<100\gev$, the
$h_1$ is mainly singlet.  As a result, the $Zh_1a_1$
coupling is very small, implying no LEP constraints on the
$h_1$ and $a_1$ from $e^+e^-\to h_1 a_1$
production.

We note that in the case of the points 1 -- 3, the
$h_2$ would not be detectable either at the LHC or
at a Linear Collider (LC). 
For points 4 -- 6, the $h_1$, though
light, is singlet in nature and would not be
detectable.  Further, the $h_3$ or $a_2$ will only
be detectable for points 1 -- 6 if a super high
energy LC is eventually built so that $e^+e^-\to
Z\to h_3 a_2$ is possible.  Thus, we will focus on
searching for the SM-like $h_1$ ($h_2$) for points
1 -- 3 (4 -- 6) using the dominant $h_1(h_2)\to
a_1a_1$ decay mode.

In the case of points 2 and 6, the $a_1\to
\tau^+\tau^-$ decays are dominant. The final state
of interest will be $jj\tau^+\tau^-$, where the
$jj$ actually comes primarily from
$a_1a_1\to\tau^+\tau^-\tau^+\tau^-$ followed by jet
decays of two of the $\tau$'s: $\tau^+\tau^-\to
jj+\nu's$.  (The contribution from direct $a_1\to jj$ decays
to the $jj\tau^+\tau^-$ final state is relatively
small for points 2 and 6.)
In what follows, when we speak of
$\tau^+\tau^-$, we refer to those $\tau$'s that
are seen in the $\tau^+\tau^-\to
\ell^+\ell^-+\nu's$ final state ($\ell=e,\mu$).
  For points 1 and 3
-- 5 $\br(a_1\to b\anti b)$ is substantial.  The
relevant final state is $b\anti b \tau^+\tau^-$.
Nonetheless, we begin with a study of the backgrounds and
signals without requiring $b$-tagging.  
With our latest cuts, 
we will see that $b$-tagging is not necessary 
to overcome the apriori large Drell-Yan
$\tau^+\tau^-$+jets background.  It is eliminated
by stringent cuts for finding the highly energetic
forward / backward
jets characteristic of the $WW$ the fusion  process.
As a result, we will find good signals for all
6 of our points.

\def\cO#1{{\cal{O}}\left(#1\right)}
\def\nn {\nonumber}
\newcommand{\bn}{\begin{enumerate}}
\newcommand{\en}{\end{enumerate}}
\newcommand{\bc}{\begin{center}}
\newcommand{\ec}{\end{center}}
\newcommand{\ul}{\underline}
\newcommand{\ol}{\overline}
\newcommand{\ar}{\rightarrow}
\newcommand{\sm}{${\cal {SM}}$}
\newcommand{\as}{\alpha_s}
\newcommand{\aem}{\alpha_{em}}
\newcommand{\ycut}{y_{\mathrm{cut}}}
\newcommand{\susy}{{{SUSY}}}
\newcommand{\Dir}{\kern -6.4pt\Big{/}}
\newcommand{\Dirin}{\kern -10.4pt\Big{/}\kern 4.4pt}
\newcommand{\DDir}{\kern -10.6pt\Big{/}}
\newcommand{\DGir}{\kern -6.0pt\Big{/}}
\def\Ecm{\ifmmode{E_{\mathrm{cm}}}\else{$E_{\mathrm{cm}}$}\fi}
\def\gluino{\ifmmode{\mathaccent"7E g}\else{$\mathaccent"7E g$}\fi}
\def\photino{\ifmmode{\mathaccent"7E \gamma}\else{$\mathaccent"7E \gamma$}\fi}
\def\mgluino{\ifmmode{m_{\mathaccent"7E g}}
             \else{$m_{\mathaccent"7E g}$}\fi}
\def\taugluino{\ifmmode{\tau_{\mathaccent"7E g}}
             \else{$\tau_{\mathaccent"7E g}$}\fi}
\def\mphotino{\ifmmode{m_{\mathaccent"7E \gamma}}
             \else{$m_{\mathaccent"7E \gamma}$}\fi}
\def\ML{\ifmmode{{\mathaccent"7E M}_L}
             \else{${\mathaccent"7E M}_L$}\fi}
\def\MR{\ifmmode{{\mathaccent"7E M}_R}
             \else{${\mathaccent"7E M}_R$}\fi}

\def\lsim{\buildrel{\scriptscriptstyle <}\over{\scriptscriptstyle\sim}}
\def\gsim{\buildrel{\scriptscriptstyle >}\over{\scriptscriptstyle\sim}}
\def\Jnl #1#2#3#4 {#1 {\bf #2} (#3) #4}
\def\NPB {{\rm Nucl. Phys.} {\bf B}}
\def\PLB {{\rm Phys. Lett.}  {\bf B}}
\def\PRL {\rm Phys. Rev. Lett.}
\def\PRD {{\rm Phys. Rev.} {\bf D}}
\def\ZPC {{\rm Z. Phys.} {\bf C}}
\def\EPJC {{\rm Eur. Phys. J.} {\bf C}}
\def\Ord{\lower .7ex\hbox{$\;\stackrel{\textstyle <}{\sim}\;$}}
\def\OOrd{\lower .7ex\hbox{$\;\stackrel{\textstyle >}{\sim}\;$}}
\def\eps{\epsilon}

In principle, one could explore final states other than 
$b\anti b \tau^+\tau^-$ (or $jj\tau^+\tau^-$ for points 2
and 6). However, all other channels will be much more
problematical at the LHC. A $4b$-signal would
be burdened by a large QCD background even after
implementing $b$-tagging.  A $4j$-signal would be
completely swamped by QCD background.  Meanwhile,
the $4\tau$-channel (by which we mean that all
taus decay leptonically) would not allow one to
reconstruct the $h_1,h_2$ resonances.  

In the case of the $2b2\tau$ (or
$2j2\tau$) signature, we identify the $\tau$'s
through their leptonic decays to electrons
and muons. Thus, they will yield some amount of
missing (transverse) momentum, $p_{\rm miss}^T$.
This missing transverse momentum can be projected
onto the visible $e,\mu$-momenta in an attempt to
reconstruct the parent $\tau$-direction.

\vspace*{-.15in}
\section{Monte Carlo Results for the LHC}

Let us now focus on the $WW\to h\to aa$
channel that we believe provides 
the best hope for Higgs detection in these
difficult NMSSM cases.  (We reemphasize that the
$h_1$ [cases 1 -- 3] or $h_2$ [cases 4 -- 6] has
nearly full SM strength coupling to $WW$.)
The $b\anti b\tau^+\tau^-$ (or
$2j\tau^+\tau^-$, for points 2 and 6) final state
of relevance is complex and subject to large
backgrounds, and the $a_1$ masses of interest are
very modest in size.  In order to extract the $WW$ fusion
$2j2\tau$ NMSSM Higgs boson signature, it is crucial
to strongly exploit forward
and backward jet tagging on the light quarks
emerging after the double $W$-strahlung preceding
$WW$-fusion.  We also require two additional central
jets (from one of the $a$'s) and two opposite sign
central leptons ($\ell=e,\mu$) coming from the 
the $\tau^+\tau^-$ emerging from the decay of the other
$a$. By imposing stringent
 forward / backward jet tagging cuts, we remove the
otherwise very large background from Drell-Yan
$\tau^+\tau^-+jets$ production. 
In the end, the most important background is due
to $t\anti t$ production and decay via the purely
SM process, $gg\to t\bar t\to b\bar b W^+W^-\to
b\bar b \tau^+\tau^- + p_{\rm miss}^T,$ in
association with forward and backward jet radiation.

\begin{figure}[h]
\begin{center}
\centerline{$~~~~~~~~~~~~~~~$LHC, $\sqrt{s_{pp}}=14$ TeV}
\centering\epsfig{file=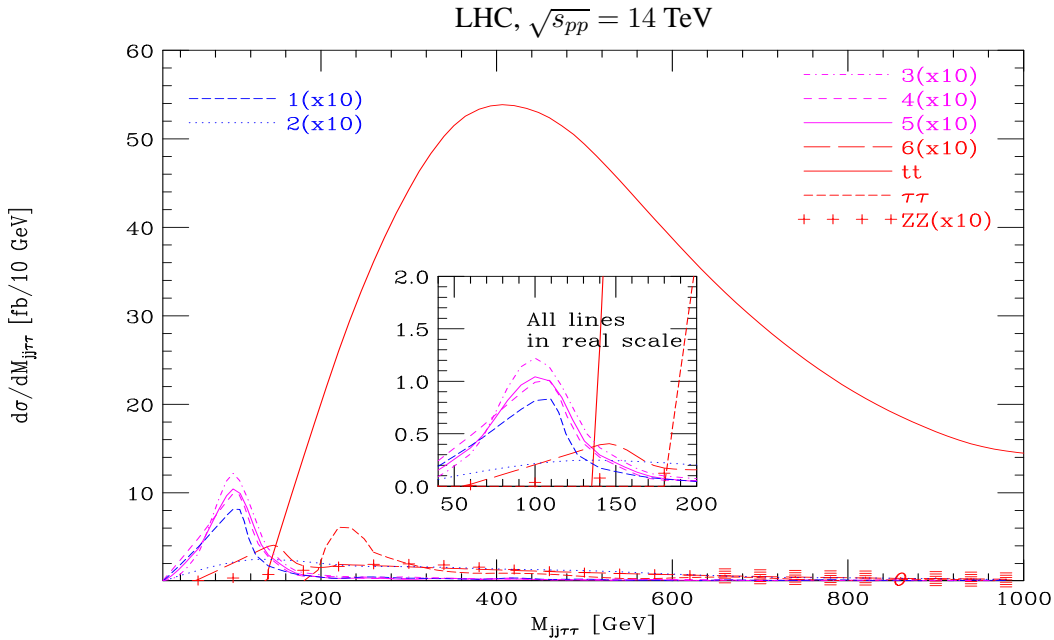,angle=90,height=8cm,width=14cm}

\vspace*{1.0truecm}

\noindent
\vspace{-1.0cm}
\caption{\footnotesize We plot $d\sigma/dM_{jj\tau^+\tau^-}$ [fb/10~GeV] vs $M_{jj\tau^+\tau^-}$~[GeV]
for signals and backgrounds after basic
event selections, but before $b$ tagging. 
The lines corresponding to points 4 and 5
are visually indistinguishable. No $K$ factors are included.
}
\label{MH}
\end{center}
\end{figure}

We have employed numerical simulations based on a version of
{\tt HERWIG v6.4}~\cite{Moretti:2002eu,Corcella:2001wc,Corcella:2000bw}
modified to allow for appropriate NMSSM couplings
and decay rates. Calorimeter emulation
was performed using the {\tt GETJET} code
\cite{GETJET}. 
Since the $a_1$ will not have been detected
  previously, we must assume a value for
  $m_{a_1}$.  In dealing with 
actual experimental data, it will be necessary to
  repeat the analysis for densely spaced $m_{a_1}$
  values and look for the $m_{a_1}$ choice that
  produces the best signal.
 We look among the central jets for the
  combination with invariant mass $M_{jj}$ closest
  to $m_{a_1}$. In Fig.~\ref{MH}, we show
the $M_{jj\tau^+\tau^-}$ invariant mass distribution
obtained after cuts, but before $b$-tagging
or inclusion of $K$ factors 
--- the plot presented assumes
that we have hit on the correct $m_{a_1}$ choice.

The selection strategy adopted is 
a more refined (as regards
forward / backward jet tagging)
version of that summarized in \cite{Ellwanger:2003jt}.
It is clearly
efficient in reconstructing the $h_1$ (for points
1--3) and $h_2$ (for points 4--6) masses from the
$jj \tau^+\tau^-$ system, as one can appreciate by
noting the peaks appearing  at
$M_{jj\tau^+\tau^-}\approx100$~GeV. In contrast,
the heavy Higgs resonances at $m_{h_2}$ for points
1--3 and the rather light resonances at $m_{h_1}$
for points 4--6 (recall Table~\ref{tpoints}) do
not appear, the former mainly because of the very
poor production rates and the latter due to the
fact that either the $h_1\to a_1 a_1$ decay mode
is not open (points 4, 5) or -- if it is -- the
jets and $e/\mu$-leptons eventually emerging
from the $a_1$ decays are too soft to pass the
acceptance cuts (point 6, for which
$m_{a_1}=7$~GeV and $m_{h_1}=51$~GeV).  For all
six NMSSM setups, the Higgs resonance produces a
bump below the end of the low mass tail of the
$t\bar t$ background (see the insert in 
Fig.~\ref{MH}).  Note how small the DY
$\tau^+\tau^-$ background is after strong
forward / backward jet tagging.  Since the main
surviving background is from $t\anti t$ production,
$b$ tagging is not helpful.  For points 2 and 6,
for which the signal has no $b$'s in the final state,
anti-$b$-tagging might be useful, but has not been
considered here.

To estimate $S/\sqrt B$, we assume $L=300~{\rm
  fb}^{-1}$, a $K$ factor of 1.1 for the $WW$
fusion signal and $K$ factors of 1, 1 and 1.6 for
the DY $\tau^+\tau^-$, $ZZ$ production and $t\anti
t$ backgrounds, respectively.  (These $K$ factors
are not included in the plot of Fig.~\ref{MH}.)
We sum events over the region $40\leq
M_{jj\tau^+\tau^-}\leq 150$~GeV.  
(Had we only included masses below $130$~GeV,
we would have had no $t\anti t$ background,
and the $S/\sqrt B$ values would be enormous.  However,
we are concerned that this absence of $t\anti t$
background below $130\gev$ might be a reflection
of limited Monte Carlo statistics.  As a result
we have taken the more conservative approach of
at least including the first few bins for which
our Monte Carlo does predict some $t\anti t$ background.)

For points 1, 2, 3,
4, 5 and 6, we obtain signal rates of about $S=1636$, 702,
  2235,  2041, 2013, and  683, respectively.
The $t\anti t$+jets background rate is $B_{tt}\sim
795$. The $ZZ$ background rate is $B_{ZZ}\sim 6$.
The DY $\tau^+\tau^-$ background rate is
negligible. (We are continuing to increase our statistics
to get a fully reliable estimate.)
The resulting $N_{SD}=S/\sqrt B$ values for points 1-6
are 50,  22, 69,  63,  62, and  21, respectively.
The smaller values for points 2 and 6 are simply
a reflection of the difficulty of
isolating and reconstructing
the two jets coming from the decay of a very light $a_1$.
Overall, these preliminary results are very encouraging
and suggest that a no-lose theorem for NMSSM Higgs
detection at the LHC is close at hand.

\vspace*{-.15in}
\section{Conclusions}

In summary, we have obtained a statistically very significant
LHC signal in the $jj\tau^+\tau^-$ final
state of $WW$ fusion for cases in which the NMSSM parameters 
are such that
the most SM-like of the CP-even Higgs bosons, $h$,
is relatively light and decays primarily to a pair
of CP-odd Higgs states, $h\to aa$ with $a\to b\anti b,\tau^+\tau^-$ if $m_a>2m_b$ or $a\to jj,\tau^+\tau^-$
if $m_a<2m_b$. 
 The statistical significances are (at least)
of order 50 to 70 for points with $m_a>2m_b$
and of order 20 for points with $m_a<2m_b$.
These high significances were obtained by
imposing stringent cuts requiring
highly energetic forward / backward jets
in order to isolate the $WW$ fusion signal process
from backgrounds such as DY $\tau^+\tau^-$ pair 
production. Still, this signal will
be the only evidence for Higgs bosons at the LHC. 
A future LC will probably be 
essential in order to confirm that
the enhancement seen at the LHC really does
correspond to a Higgs boson. At the LC, discovery
of a light SM-like $h$ is guaranteed to be
possible in the $Zh$ final state using the recoil
mass technique~\cite{Gunion:2003fd}.  

In the present study, we have not
explored the cases in which the
$a_1\to\cnone\cnone$ decay has a large branching
ratio.  Detecting a Higgs signal in such cases
will require a rather different procedure.
Work on the $WW\to h\to {\rm invisible}$ signal
is in progress~\cite{preparation}.

As we have stressed, for parameter space points of
the type we have discussed here, detection of any
of the other MSSM Higgs bosons is likely to be
impossible at the LHC and is likely to require an
LC with $\sqrt{s_{e^+e^-}}$ above the relevant
thresholds for $h'a'$ production, where $h'$ and
$a'$ are heavy CP-even and CP-odd Higgs bosons,
respectively.  

Although results for the LHC
indicate that Higgs boson discovery will be possible
for the type of situations we have
considered, it is clearly important
to refine and improve the techniques for extracting a
signal. This will almost certainly be possible 
once data is in
hand and the $t\anti t$ background can be more
completely modeled.  

Clearly, if SUSY is
discovered and $WW\to WW$ scattering is found to
be perturbative at $WW$ energies of 1 TeV (and
higher), and yet no Higgs bosons are detected in
the standard MSSM modes, a careful search for the
signal we have considered should have a high
priority.  

Finally, we should remark that the
$h\to aa$ search channel considered here in the
NMSSM framework is also highly relevant for a
general two-Higgs-doublet model, 2HDM.  It is
really quite possible that the most SM-like
CP-even Higgs boson of a 2HDM will decay primarily
to two CP-odd states.  This is possible even if
the CP-even state is quite heavy, unlike the NMSSM
cases considered here. If CP violation is introduced
in the Higgs sector, either at tree-level
or as a result of one-loop corrections (as, for example,
is possible in the MSSM),
$h\to h' h''$ decays will generally be present. The 
critical signal will be the same as
that considered here.

\subsubsection*{Acknowledgments}
JFG is supported by the U.S. Department of Energy and the Davis
Institute for High Energy Physics. SM thanks the UK-PPARC 
and the Royal Society (London, U.K.) for financial support
and D.J. Miller for useful conversations. 
CH is supported by the European Commission RTN grant HPRN-CT-2000-00148.
JFG, CH, and UE thank
the France-Berkeley fund for partial support of this research.

%\begin{thebibliography}{99}
%\input{nmssm_leshouches_proc_bib.tex}
%\end{thebibliography}

\bibliography{nmssm_leshouches_proc}

\end{document}